# Persistent hot carrier diffusion in boron arsenide single crystals imaged by ultrafast electron microscopy

Usama Choudhry[1]†, Fengjiao Pan[2]†, Taeyong Kim[1], Ryan Gnabasik[1], Geethal Amila Gamage[2], Haoran Sun[2], Alex Ackerman[1], Zhifeng Ren[2]*, Bolin Liao[1]*

[1]Department of Mechanical Engineering, University of California Santa Barbara; Santa Barbara, California 93110, USA.

[2]Department of Physics and Texas Center for Superconductivity, University of Houston; Houston, Texas 77204 USA.

†These authors contributed equally to this work

*Corresponding author. Email: zren2@central.uh.edu; bliao@ucsb.edu

**Abstract:** Cubic boron arsenide (BAs) is promising for microelectronics thermal management due to its high thermal conductivity. Recently, its potential as an optoelectronic material is also being explored. However, it remains challenging to measure its photocarrier transport properties due to small sizes of available high-quality crystals. Here, we use scanning ultrafast electron microscopy (SUEM) to directly visualize the diffusion of photoexcited charge carriers in BAs single crystals. Surprisingly, we observed ambipolar diffusion at low optical fluence with persistent hot carrier dynamics for above 200 picoseconds, which can be attributed to the large frequency gap between acoustic and optical phonons, the same feature that is responsible for the high thermal conductivity. At higher optical fluence, we observed spontaneous electron-hole separation. Our results show BAs is an attractive optoelectronic material combining high thermal conductivity and excellent photocarrier transport properties. Our study also demonstrates the capability of SUEM to probe photocarrier transport in emerging materials.

**Main Text:** Historically, BAs is the least studied semiconductor in the III-V family, likely due to the difficulty in growing high quality crystals[1]. Recent theoretical calculations[2,3] and experimental measurements[4–6] have confirmed that BAs possesses a high thermal conductivity above 1000 W/mK near room temperature. This unexpected value is caused by unique features in the phonon dispersion in BAs that limit the possible channels of phonon-phonon scatterings, including the large acoustic-optical phonon band gap and the closely "bunched" acoustic branches[2]. Furthermore, BAs has a thermal expansion coefficient that matches that of silicon and other III-V semiconductors, making it a suitable substrate material for device thermal management[7]. Besides the excellent thermal transport properties, BAs is a semiconductor with a moderate band gap of 2 eV[8]. First-principles simulation[9] has suggested a very high intrinsic charge mobility limited by electron-phonon interaction in BAs (electron: 1400 $cm^2$/Vs; hole: 2110 $cm^2$/Vs). This is attributed to suppressed polar optical phonon scattering of charge carriers[9]. In addition, slow cooling of photoexcited hot carriers has been predicted in BAs using first-principles simulation, thanks to the large acoustic-optical phonon band gap that limits the thermalization process within the phonon system[10]. If experimentally verified, the combination of high thermal conductivity, high charge mobility, and prolonged hot carrier dynamics will make BAs an extraordinary candidate material for future optoelectronic, photovoltaic[11] and photocatalytic applications[12]. However, measuring charge transport properties in BAs has been challenging due to the lack of high-quality single crystals with a sufficient size for conventional bulk measurements, such as Hall effect and field effect studies. Previous electrical transport experiments have suggested a hole mobility below 400 $cm^2$/Vs,[13,14] most likely due to the high concentration of impurities[15]. Moreover, there is currently no experimental report of the hot photocarrier dynamics in BAs.

In this work, we report time-resolved imaging of photoexcited charge transport in high-quality BAs single crystals using SUEM. SUEM is a photon-pump-electron-probe technique that uses short electron pulses with picosecond duration to image the response of a sample surface after the impact of a photon pulse[16,17]. It integrates the temporal resolution of femtosecond lasers with the spatial resolution of scanning electron microscopes (SEMs). The change of local secondary electron (SE) yield as a result of the optical excitation is measured and used to form contrast images[18]. Given the shallow escape length of SEs (a few nanometers), SUEM is highly sensitive to surface charge dynamics, and has been used to study photocarrier diffusion in uniform materials[19–21] and near interfaces[22,23]. A schematic of our experiment is shown in fig. 1A. Details can be found in Methods. Briefly, we split the output from a femtosecond pulsed laser (wavelength: 1030 nm, pulse duration: 150 fs, repetition rate: 5 MHz) into a pump beam and a photoelectron-generation beam. The pump beam is converted to 515 nm, passed through a mechanical delay stage, and focused onto the sample to excite photocarriers. The photoelectron-generation beam is converted to 257 nm and focused onto the Schottky electron gun to generate electron pulses as the probe (the "primary electrons", PEs). The relative delay between the optical pump pulses and the PE probe pulses is controlled by the delay stage. The PE probe pulses are accelerated to 30 keV, focused and scanned over the sample surface by electron optics inside the SEM column. SEs emitted from each location on the sample surface as a result of the PE impact are collected by an Everhart-Thornley detector (ETD). The change of local SE yield after optical excitation, which is sensitive to the presence of photoexcited electrons and holes near the sample surface[17], is used to form SUEM contrast images. These images directly visualize the surface distribution of photocarriers at various delay times after optical excitation. The spatial resolution of these images is determined by the size of the PE beam (a few nanometers) and the time resolution is limited by the duration of the PE pulses (a few picoseconds with less than 100 electrons per pulse)[24,25].

Our BAs samples are prepared using a chemical vapor transport method, as reported elsewhere[26]. Details can be found in Methods. The SEM image of a representative sample is shown in fig. 1B. The x-ray diffraction (XRD) pattern shown in fig. 1C verifies the high quality of the single crystal samples measured in this study. SEM-based energy dispersive spectroscopy (EDS) (fig. 1D) suggests that carbon and oxygen are the major impurities. Raman spectrum is shown in fig. 1E, where the narrow linewidth suggests a low impurity concentration.

Figure 2A displays representative SUEM contrast images obtained on a BAs single crystal. For this dataset, an optical pump fluence of 100 $\mu J/cm^2$ was used. More datasets taken from different samples with different optical pump fluences are included in the Supplementary Materials (table S1, and figs. S1-S5). Each contrast image was obtained by subtracting a reference SE image taken at -700 ps delay time from the SE image taken at a later delay time. Thus, the SUEM contrast images reflect the change in SE emission from the sample surface as a result of the pulsed photoexcitation that arrives at time zero. As shown in fig. 2A, no obvious contrast was observed before time zero, indicating that the time interval between pump pulses (200 ns at 5 MHz repetition rate) is sufficiently long for the excited sample to return to the equilibrium state. After time zero, a bright contrast within the pump-illuminated region emerges and its size grows in time. The bright contrast signals a higher SE yield from the region, which is caused by the presence of photoexcited charge carriers. Photocarriers modify the average energy of the electrons locally that can affect the SE yield[17]. In addition, photocarriers can modulate the surface electrostatic potential (i.e., the surface photovoltage effect[18]), providing another mechanism for the SE yield contrast. Accordingly, we interpret the bright contrast in the SUEM images as an indicator for the distribution of the photocarriers near the sample surface. Therefore, the growth in the spatial size of the bright contrast as a function of the delay time between the optical pump and the electron probe corresponds to the lateral diffusion of photocarriers. In this manner, the dynamics of photocarriers near the sample surface can be directly visualized in the SUEM contrast images. When the optical pump fluence is low, photogenerated electrons and holes diffuse together with an intermediate diffusivity due to their Coulombic interaction. This is called the "ambipolar diffusion" regime. From fig. 2A, we observe an initial fast expansion of the bright contrast up to a few hundred picoseconds when the expansion starts to slow down.

Next, we quantitatively analyze the diffusion process visualized in the SUEM contrast images. Since the optical pump beam has a Gaussian spatial profile with a $1/e^2$ radius of 30 $\mu$m, the initial distribution of the photocarrier concentration should also follow a Gaussian profile in the radial direction. Along the sample thickness direction, the optical pump is absorbed with a penetration depth on the order of a few microns[8]. A numerical simulation suggests that the diffusion along the thickness direction does not notably change the dynamics of the lateral radius of the photocarrier distribution near the sample surface. Since the lateral diffusion process preserves the Gaussian spatial profile, we fit the spatial distribution of the bright contrast at each delay time to a two-dimensional Gaussian function with a varying radius using a least-square fitting algorithm (some representative fits are shown in fig. 2B). The fitted squared radius of the photocarrier distribution as a function of delay time is shown in fig. 3A. The error bars represent the 95% confidence intervals. Early data points (< 40 ps) are noisy due to the initial weak contrast, and thus, are not shown here for clarity (they are included in datasets in the Supplementary Materials). Consistent with the SUEM contrast images shown in fig. 2A, the fitted squared radius shows an initial fast diffusion regime that slows down after a few hundred picoseconds. The initial fast diffusion was also observed in crystalline silicon[19] and amorphous silicon[20] and is a result of the transport of photoexcited hot carriers. Since the photon energy of the optical pump ($h\nu_p = 2.4$ eV) is higher than the band gap in BAs ($E_G \approx 2$ eV)[8], the initial temperature of the photocarriers after excitation can be estimated to be $T_i = (h\nu_p - E_G)/k_B \approx 4700$ K. In addition, the hot photocarriers occupy higher energy states from the band edges and can possess higher band velocities and experience reduced electron-phonon scattering[27]. Therefore, the effective diffusivity of the hot photocarriers can be orders of magnitude higher than the equilibrium value[19]. Due to electron-phonon scattering, however, the hot photocarriers will cool down to the band edges and the diffusivity will approach the equilibrium value. To quantitatively model this process, we assume the effective diffusivity of the photocarriers decays exponentially with a time constant $\tau$: $D(t) = (D_i - D_0)e^{-t/\tau} + D_0$, where $D_i$ is the effective diffusivity immediately after photoexcitation, and $D_0$ is the equilibrium diffusivity at 300 K. With this time-dependent diffusivity, we solved the two-dimensional diffusion equation analytically (Supplementary Materials) and compared the solution (eqn. S-4) to the experimental data, as shown in fig. 3A. We used the effective equilibrium diffusivity $D_0 \approx 45$ cm$^2$/s, corresponding to an equilibrium ambipolar mobility $\mu \approx 1700$ cm$^2$/Vs, based on the mobility calculated from first principles[9] and the Einstein relation, and kept the time constant $\tau$ as a fitting parameter. For the dataset shown in fig. 2 and fig. 3A, the fitted time constant $\tau$ is 220 ps. To show the model sensitivity to these parameters, we also plot bounds corresponding to varying $\mu$ and $\tau$ by $\pm 20\%$ in fig. 3A. More datasets are presented in the Supplementary Materials. Within these datasets, the time constant $\tau$ ranges from 145 ps to 250 ps and seems to decrease with increasing optical fluence, suggesting the onset of Auger recombination. A dataset with higher optical fluence (130 $\mu$J/cm$^2$) and a fitted time constant $\tau$ of 160 ps is shown in fig. 3B.

The hot carrier transport time constant $\tau$ (up to 250 ps) extracted from our SUEM measurements of BAs is significantly longer than that in Si (~100 ps on average)[19]. The slow hot carrier cooling process is likely due to the same feature in the phonon dispersion that leads to the high thermal conductivity in BAs: the large acoustic-optical phonon band gap caused by the large mass ratio between boron and arsenic atoms[10]. Owing to the strong polar optical phonon scattering of electrons in BAs, the hot carrier energy is first transferred to the polar optical phonons, and the large acoustic-optical phonon frequency gap prevents effective scattering between optical and acoustic phonons, creating a bottleneck in the hot carrier cooling pathway. The prolonged hot carrier transport can be an attractive feature for photovoltaic[11] and photocatalytic applications[12].

We conducted SUEM measurements using different optical pump fluences. When the optical fluence stayed below 130 $\mu$J/cm$^2$, we did not observe any qualitative change in the photocarrier dynamics (Supplementary Materials), suggesting that the observed dynamics were within the linear regime. At a higher fluence above 180 $\mu$J/cm$^2$, however, we observed qualitatively distinct features in the SUEM contrast images: the formation of an outer dark region and an inner bright region, both regions expanding with time, as shown in fig. 4A. The optical fluence for this dataset is 235 $\mu$J/cm$^2$. Another dataset with 185 $\mu$J/cm$^2$ fluence is shown in fig. S5. This observation can be explained by separate diffusion of faster holes and slower electrons[9] beyond the ambipolar diffusion regime (fig. 4B). This phenomenon is reminiscent of the photo-Dember effect[28] and similar to the charge separation observed in silicon, where electrons diffuse faster than holes, under higher optical pump fluences[29]. Figure 4C shows the evolution of the areas of the dark and the bright regions as a function of the delay time, where the initial separation and the convergence towards ambipolar diffusion can be observed. We used eqn. S-4 with the calculated equilibrium electron and hole mobilities to fit the experimental data and obtained time constants 256 ps for electrons and 767 ps for holes, as shown in fig. 4C. However, we caution here that a detailed coupled electron-hole transport model incorporating their Coulombic interactions is needed for a rigorous analysis. A recent SUEM study suggests that vertical photocarrier transport driven by a surface field can also lead to fast lateral expansion of the SUEM contrast[30]. In contrast, the lack of contrast saturation at high optical fluences in our study suggests that the vertical transport does not play a significant role here. In addition, the vertical transport picture cannot explain the complex contrast features we observed at higher fluences.

Our study reveals that BAs has excellent photocarrier transport properties, in addition to its high thermal conductivity. These unusual properties are attributed to its unique electron and phonon band structures, especially the large acoustic-optical phonon frequency gap. The combination of desirable optoelectronic and thermal properties of BAs renders it an exciting semiconducting material. We also demonstrated SUEM as an emerging tool to directly visualize photocarrier dynamics in samples that are otherwise difficult to characterize with conventional methods.

**References and Notes**
1. Perri, J. A., La Placa, S. & Post, B. New group III-group V compounds: BP and BAs. *Acta Cryst* **11**, 310–310 (1958).
2. Lindsay, L., Broido, D. A. & Reinecke, T. L. First-Principles Determination of Ultrahigh Thermal Conductivity of Boron Arsenide: A Competitor for Diamond? *Phys. Rev. Lett.* **111**, 025901 (2013).
3. Feng, T., Lindsay, L. & Ruan, X. Four-phonon scattering significantly reduces intrinsic thermal conductivity of solids. *Phys. Rev. B* **96**, 161201 (2017).

4. Tian, F. *et al.* Unusual high thermal conductivity in boron arsenide bulk crystals. *Science* **361**, 582–585 (2018).

5. Kang, J. S., Li, M., Wu, H., Nguyen, H. & Hu, Y. Experimental observation of high thermal conductivity in boron arsenide. *Science* **361**, 575–578 (2018).

6. Li, S. *et al.* High thermal conductivity in cubic boron arsenide crystals. *Science* **361**, 579–581 (2018).

7. Kang, J. S. *et al.* Integration of boron arsenide cooling substrates into gallium nitride devices. *Nat Electron* **4**, 416–423 (2021).

8. Song, B. *et al.* Optical properties of cubic boron arsenide. *Appl. Phys. Lett.* **116**, 141903 (2020).

9. Liu, T.-H. *et al.* Simultaneously high electron and hole mobilities in cubic boron-V compounds: BP, BAs, and BSb. *Phys. Rev. B* **98**, 081203 (2018).

10. Sadasivam, S., Chan, M. K. Y. & Darancet, P. Theory of Thermal Relaxation of Electrons in Semiconductors. *Phys. Rev. Lett.* **119**, 136602 (2017).

11. König, D. *et al.* Hot carrier solar cells: Principles, materials and design. *Physica E: Low-dimensional Systems and Nanostructures* **42**, 2862–2866 (2010).

12. Zhou, L. *et al.* Quantifying hot carrier and thermal contributions in plasmonic photocatalysis. *Science* **362**, 69–72 (2018).

13. Kim, J. *et al.* Thermal and thermoelectric transport measurements of an individual boron arsenide microstructure. *Appl. Phys. Lett.* **108**, 201905 (2016).

14. Chu, T. L. & Hyslop, A. E. Crystal Growth and Properties of Boron Monoarsenide. *Journal of Applied Physics* **43**, 276–279 (1972).

15. Chen, X. *et al.* Effects of Impurities on the Thermal and Electrical Transport Properties of Cubic Boron Arsenide. *Chem. Mater.* **33**, 6974–6982 (2021).

16. Yang, D.-S., Mohammed, O. F. & Zewail, A. H. Scanning ultrafast electron microscopy. *Proceedings of the National Academy of Sciences* **107**, 14993–14998 (2010).

17. Liao, B. & Najafi, E. Scanning ultrafast electron microscopy: A novel technique to probe photocarrier dynamics with high spatial and temporal resolutions. *Materials Today Physics* **2**, 46–53 (2017).

18. Li, Y., Choudhry, U., Ranasinghe, J., Ackerman, A. & Liao, B. Probing Surface Photovoltage Effect Using Photoassisted Secondary Electron Emission. *J. Phys. Chem. A* **124**, 5246–5252 (2020).

19. Najafi, E., Ivanov, V., Zewail, A. & Bernardi, M. Super-diffusion of excited carriers in semiconductors. *Nature Communications* **8**, 15177 (2017).

20. Liao, B., Najafi, E., Li, H., Minnich, A. J. & Zewail, A. H. Photo-excited hot carrier dynamics in hydrogenated amorphous silicon imaged by 4D electron microscopy. *Nature Nanotechnology* **12**, 871 (2017).

21. Liao, B. *et al.* Spatial-Temporal Imaging of Anisotropic Photocarrier Dynamics in Black Phosphorus. *Nano Lett.* **17**, 3675–3680 (2017).

22. Najafi, E., Scarborough, T. D., Tang, J. & Zewail, A. Four-dimensional imaging of carrier interface dynamics in p-n junctions. *Science* **347**, 164–167 (2015).

23. Wong, J. *et al.* Spatiotemporal Imaging of Thickness-Induced Band-Bending Junctions. *Nano Lett.* **21**, 5745–5753 (2021).

24. Mohammed, O. F., Yang, D.-S., Pal, S. K. & Zewail, A. H. 4D Scanning Ultrafast Electron Microscopy: Visualization of Materials Surface Dynamics. *J. Am. Chem. Soc.* **133**, 7708–7711 (2011).

25. Sun, J., Melnikov, V. A., Khan, J. I. & Mohammed, O. F. Real-Space Imaging of Carrier Dynamics of Materials Surfaces by Second-Generation Four-Dimensional Scanning Ultrafast Electron Microscopy. *J. Phys. Chem. Lett.* **6**, 3884–3890 (2015).

26. Gamage, G. A., Chen, K., Chen, G., Tian, F. & Ren, Z. Effect of nucleation sites on the growth and quality of single-crystal boron arsenide. *Materials Today Physics* **11**, 100160 (2019).

27. Sjodin, T., Petek, H. & Dai, H.-L. Ultrafast Carrier Dynamics in Silicon: A Two-Color Transient Reflection Grating Study on a (111) Surface. *Phys. Rev. Lett.* **81**, 5664–5667 (1998).

28. Klatt, G. *et al.* Terahertz emission from lateral photo-Dember currents. *Opt. Express, OE* **18**, 4939–4947 (2010).

29. Najafi, E., Jafari, A. & Liao, B. Carrier density oscillation in the photoexcited semiconductor. *J. Phys. D: Appl. Phys.* **54**, 125102 (2021).

30. Ellis, S. R. *et al.* Scanning ultrafast electron microscopy reveals photovoltage dynamics at a deeply buried p-Si/SiO$_2$ interface. *Phys. Rev. B* **104**, L161303 (2021).

31. Hadjiev, V. G., Iliev, M. N., Lv, B., Ren, Z. F. & Chu, C. W. Anomalous vibrational properties of cubic boron arsenide. *Phys. Rev. B* **89**, 024308 (2014).

32. Sun, H. *et al.* Boron isotope effect on the thermal conductivity of boron arsenide single crystals. *Materials Today Physics* **11**, 100169 (2019).

**Acknowledgments:**

**Funding:** The work conducted at University of California Santa Barbara is based on research supported by US Department of Energy, Office of Basic Energy Sciences, under the award number DE-SC0019244 (for the development of SUEM) and by the US Army Research Office under the award number W911NF-19-1-0060 (for studying photocarrier dynamics in emerging materials). The growth of high quality BAs crystals at University of Houston was supported by the US Office of Naval Research under Multidisciplinary University Research Initiative grant N00014-16-1-2436.

**Author contributions:**

This project was conceptualized and supervised by ZR and BL. Samples was prepared by FP, GG, HS and ZR. FP carried out the XRD and Raman characterization. SUEM was developed by UC, TK, RG, AA and BL. SUEM measurements were done by UC. SUEM data was analyzed by UC and BL. The manuscript was drafted by UC, FP, BL and ZR, and was edited with input from all authors.

**Competing interests:** US patent publication no. 20210269318 has been granted to the method for growing BAs crystals in this work.

**Data and materials availability:** All data are available in the main text or the supplementary materials.

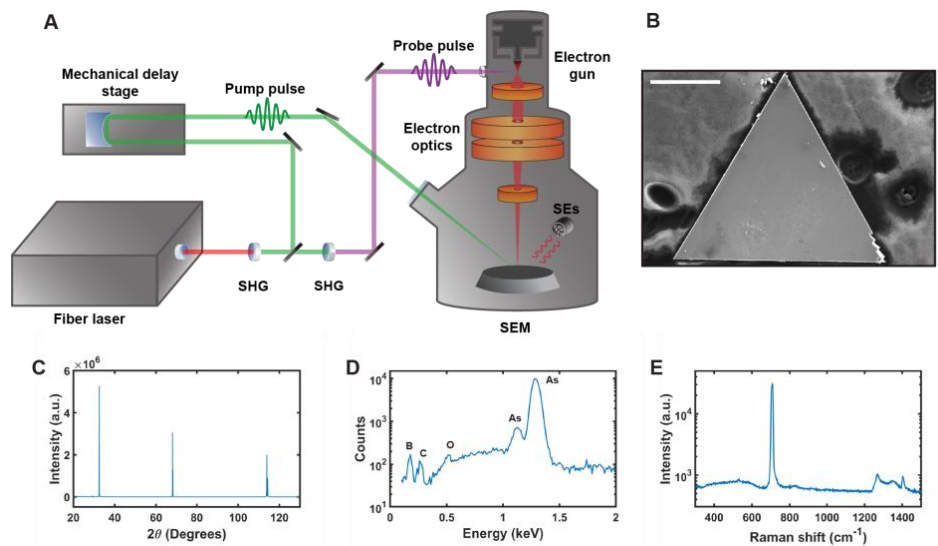

**Fig. 1. Schematic of SUEM experiment and characterization of BAs samples.** (**A**) Schematic of the SUEM experiment. SHG: second-harmonic generation. SEs: secondary electrons. (**B**) SEM image of a representative BAs flake. Scale bar: 200 $\mu$m. (**C**) XRD pattern of a representative BAs sample. (**D**) EDS spectrum of a BAs flake. (**E**) Raman spectrum of a BAs flake.

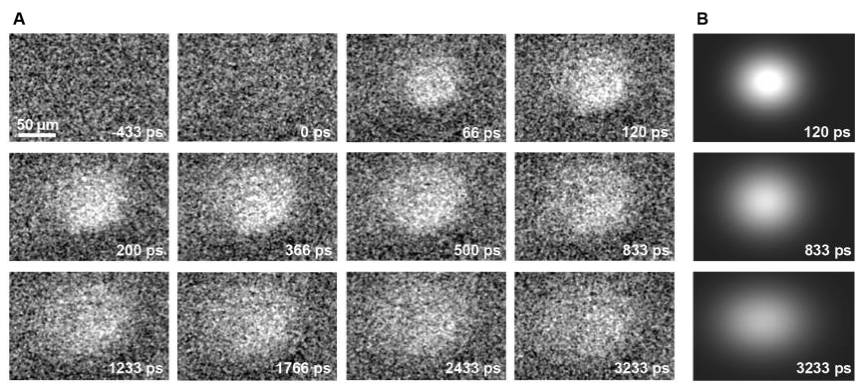

**Fig. 2. SUEM difference images from BAs crystals.** (**A**) SUEM difference images from selected time points throughout the experiment. No contrast is visible at negative time, before the arrival of the pump pulse (at 0 ps). After the arrival of the pump pulse, a small area of bright contrast emerges. This region rapidly expands for a few hundred picoseconds, before continuing to slowly expand and decrease in intensity over the next few nanoseconds. (**B**) Representative two-dimensional Gaussian fits of the experimental contrast images.

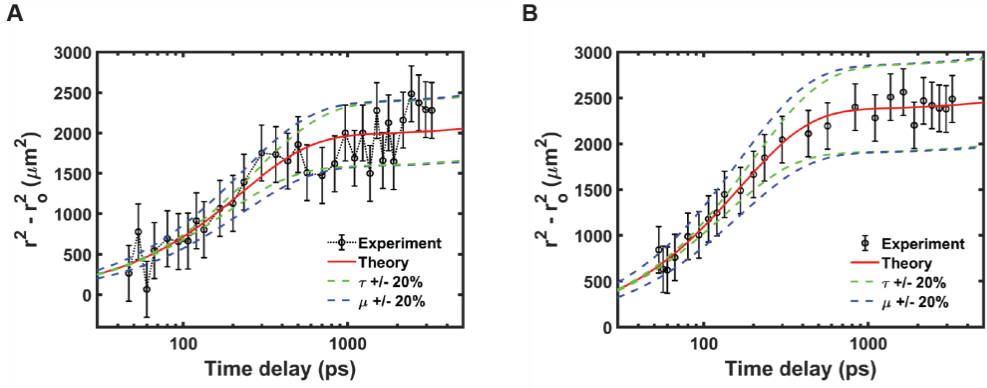

**Fig. 3. Evolution of the bright contrast in the SUEM images.** (**A**) Squared radius ($r^2$) of the bright contrast as a function of the time delay with an optical fluence of 100 $\mu J/cm^2$. (**B**) Squared radius of the bright contrast as a function of the time delay with an optical fluence of 130 $\mu J/cm^2$. $r_0$ is the radius of the optical pump beam that defines the radius of the initial photocarrier concentration distribution. The black circles label the experimental data. The dotted black line serves as eye guide. The red solid line denotes the theoretical model. The green and blue dashed lines label the theoretical model when the time constant $\tau$ and the equilibrium ambipolar mobility $\mu$ by $\pm 20\%$.

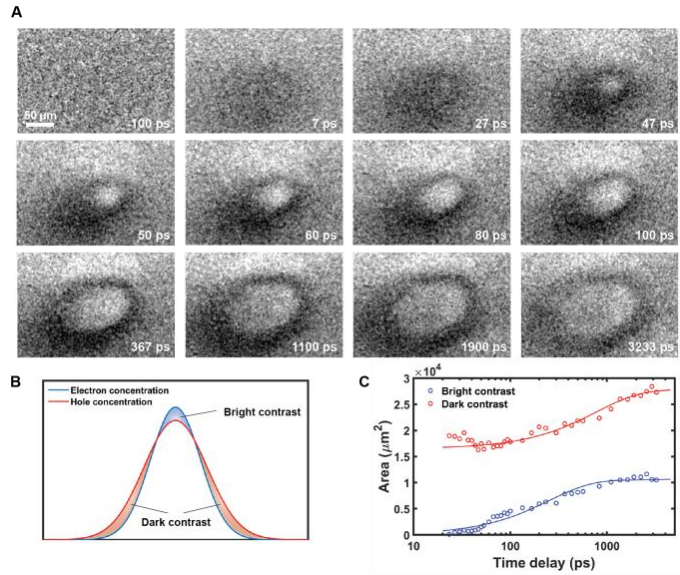

**Fig. 4. SUEM images at high optical fluence showing separation of electrons and holes.** (**A**) The SUEM contrast images taken with an optical fluence of 235 $\mu J/cm^2$. (**B**) A schematic showing the separation of electrons and holes and the corresponding SUEM contrasts. (**C**) The evolution of the areas of the bright and the dark contrast regions as a function of delay time. The solid lines represent fits using eqn. S-4 with parameters given in the main text.

**Methods**

Scanning ultrafast electron microscopy

This section provides detailed description on the SUEM setup employed in this work. A fundamental infrared (IR) laser pulse train (Clark MXR IMPULSE, wavelength: 1030 nm, pulse duration: 150 fs, repetition rate: 5 MHz) is directed to frequency-doubling crystals to create the visible pump beam (wavelength: 515 nm with tunable power) and the ultraviolet (UV) photoelectron excitation beam (wavelength: 343 nm, power: 15 mW). The visible pump beam travels variable distances adjusted by a mechanical delay stage (Newport DL600, delay time range: -0.7 ns to 3.3 ns). The UV excitation beam is directed through a transparent window on the column of an SEM (ThermoFisher Quanta 650 FEG) and onto the apex of a cooled Schottky field emission gun (a zirconium-oxide-coated tungsten tip), generating electron pulses with sub-picosecond durations via the photoelectric effect. An electron current (I) of 25 to 35 pA is used in this experiment, corresponding to 30 to 40 electrons per pulse. A time resolution of 3 ps is expected for 40 electrons per pulse[25]. The photo-generated electron pulses are accelerated inside the SEM column to 30 keV kinetic energy, and are finely focused to nanometer size through the electron optics in the SEM. The beam dwell time at each pixel was 300 ns. Each SUEM contrast image at a given delay time represented an average of 2000 to 4000 images. During the measurements, the photocathode was refreshed every 60 minutes to prevent the fluctuation of the cathode work function. A mechanical coupling system is built to make a rigid connection between the SEM air-suspension system and the optical table hosting the laser and the optical system to minimize the relative vibration that affects the alignment at the photocathode.

Sample preparation

Cubic boron arsenide (BAs) single crystals were synthesized using a chemical vapor transport (CVT) method. First, natural boron (B) (19.9% $^{10}$B and 80.1% $^{11}$B, 99.9+% purity, UMC), pure arsenic (As, 99.99999+%, Alfa Aesar), together with appropriate amount of transport agent iodine ($I_2$, 99.9985+%, Alfa Aesar), were sealed in a fused quartz tube under vacuum ($10^{-4}$ Torr). The sealed-quartz tube was then put into a two-zone horizontal tube furnace. All materials were placed at the source end of the sealed quartz tube, which was positioned at the high-temperature zone of 890 ℃. The other end of the quartz tube was positioned at the low-temperature zone of 790 ℃. At the source end, $I_2$ reacted with B to form gaseous species BI and $BI_3$ and transported to the growth end, and finally reacted with As to form BAs. After three weeks, BAs crystal pieces were obtained from the quartz tube.

XRD measurement

X-ray diffraction (XRD) measurement was conducted by a Riguka SmartLab X-ray diffractometer with a Cu K$\alpha$ radiation source. High-quality BAs crystals are usually reddish plates, with the two largest surfaces usually parallel to each other. For the measurement, the BAs sample was placed on a zero-background holder with one of the largest surfaces attached to the holder. Figure 1C shows the XRD pattern of a BAs sample with $2\theta$ ranging from 20 ° to 120 °. Each peak in this pattern includes two peaks from K$\alpha_1$ and K$\alpha_2$. The XRD measurement showed that the BAs samples grown by the CVT method are high-quality single crystals, with the preferred orientation of (111).

Raman measurement

Raman spectra were measured by SpectraPro HRS-300 (Princeton Instruments) using 1200 g/mm grating. 532 nm laser was used to excite the sample, and the laser intensity is ~6 mW. Figure 1E shows the Raman spectra of a point on the c-BAs sample. As has been reported previously[31,32], BAs has only one phonon that is active in the first-order Raman scattering, since the LO-TO splitting is very small. To improve the resolution of the Raman spectra, the slit was adjusted to a minimum to get the Raman spectra of the sharp LO peak at ~705 cm$^{-1}$. The full-width at half maximum (FWHM) of the LO peak has a small value of ~5.3 cm$^{-1}$, which indicates low impurity concentrations.

Energy dispersive spectroscopy (EDS) measurement

EDS measurement was conducted within the same SEM for the SUEM measurement (ThermoFisher Quanta 650 FEG). 10 keV electron energy was used for the measurement.

# Supplementary Materials for

## Persistent hot carrier diffusion in boron arsenide single crystals imaged by ultrafast electron microscopy

Usama Choudhry, Fengjiao Pan, Taeyong Kim, Ryan Gnabasik, Geethal Amila Gamage, Haoran Sun, Alex Ackerman, Zhifeng Ren, Bolin Liao

Correspondence to: zren2@central.uh.edu; bliao@ucsb.edu

**This PDF file includes:**

    Supplementary Text
    Table S1
    Figs. S1 to S5
    Captions for Movies S1 to S7

**Other Supplementary Materials for this manuscript include the following:**

    Movies S1 to S7

**Supplementary Text**

Diffusion model

We used a simple diffusion model to analyze our experimental data. Given the axial symmetry of our experimental geometry, the diffusion process of photocarriers is governed by the following diffusion equation:

$$\frac{\partial n}{\partial t} = D(t)\left[\frac{1}{r}\frac{\partial}{\partial r}\left(r\frac{\partial n}{\partial r}\right) + \frac{\partial^2 n}{\partial z^2}\right], \tag{S-1}$$

where $n$ is the density of electrons or holes, $D(t)$ is a time-dependent effective diffusivity, $r$ is the radial distance from the center of the optically excited area, $z$ is the depth into the sample from the surface. We assumed the effective diffusivity $D(t)$ decays exponentially in time as the hot carriers cool down, as explained in the main text. Immediately after the photoexcitation, the distribution of the photocarrier density has the following form:

$$n(r, t=0) = n_0 e^{-2r^2/R_0^2} e^{-z/d}, \tag{S-2}$$

where $n_0$ is the photocarrier density at the center of the illuminated area on the surface, $R_0$ is the $1/e^2$ radius of the optical pump beam, and $d$ is the optical absorption depth. There is large uncertainty in the optical absorption depth in BAs as determined by different methods, and the reported optical absorption depth in BAs varies from 1 $\mu$m to tens of micrometers(*8*). We used a finite differencing method to solve eqn. (S-1) numerically. We found that the diffusion along the thickness (*z*) direction changes the magnitude of the surface photocarrier density appreciably but has a small impact on the temporal evolution of the radius of the surface photocarrier density distribution within the range of possible optical absorption depth. Therefore, to understand the dynamics of the surface photocarriers as observed in our experiment, we also analytically solved a two-dimensional radial diffusion equation without the *z* direction diffusion term. By assuming that the surface photocarrier density distribution follows a Gaussian distribution with a time-dependent radius: $n(r,t) = n_0(t) e^{-2r^2/R^2(t)}$, substitution of this solution into the radial diffusion equation leads to the following equation governing the time dependence of $R(t)$:

$$\frac{dR}{dt} = \frac{2D(t)}{R} = \frac{2}{R}\left[(D_i - D_0)e^{-t/\tau} + D_0\right]. \tag{S-3}$$

This equation can be analytically solved to give the following time dependence of $R(t)$:

$$R^2 - R_0^2 = 4(D_i - D_0)\tau(1 - e^{-t/\tau}) + 4D_0 t. \tag{S-4}$$

Given the large uncertainty in the optical absorption depth $d$ and the weak dependence of $R(t)$ on the $z$-direction diffusion, we used eqn. (S-4) to fit our experimental models and good agreement was achieved.

| Designation | Sample | Measurement Date | Optical Fluence (µJ/cm$^2$) | τ (ps) | Shown in Figure | Movie |
|---|---|---|---|---|---|---|
| s1-m1 | Flake 1 | Jan. 13, 2022 | 100 | 220 | Fig. 2A<br>Fig. 3A | S1 |
| s1-m2 | Flake 1 | Feb. 3, 2022 | 80 | 233 | Fig. S1 | S2 |
| s1-m3 | Flake 1 | Feb. 4, 2022 | 130 | 145 | Fig. S2 | S3 |
| s2-m1 | Flake 2 | Feb. 5, 2022 | 130 | 160 | Fig. 3B<br>Fig. S3 | S4 |
| s2-m2 | Flake 2 | Feb. 6, 2022 | 80 | 250 | Fig. S4 | S5 |
| hf-m1 | Flake 1 | Jan. 17, 2022 | 235 | | Fig. 4 | S6 |
| hf-m2 | Flake 2 | Jan. 16, 2022 | 185 | | Fig. S5 | S7 |

**Table S1. Summary of the SUEM measurements.** The first 5 datasets were obtained using lower optical fluences, while the last 2 datasets were obtained using higher optical fluences.

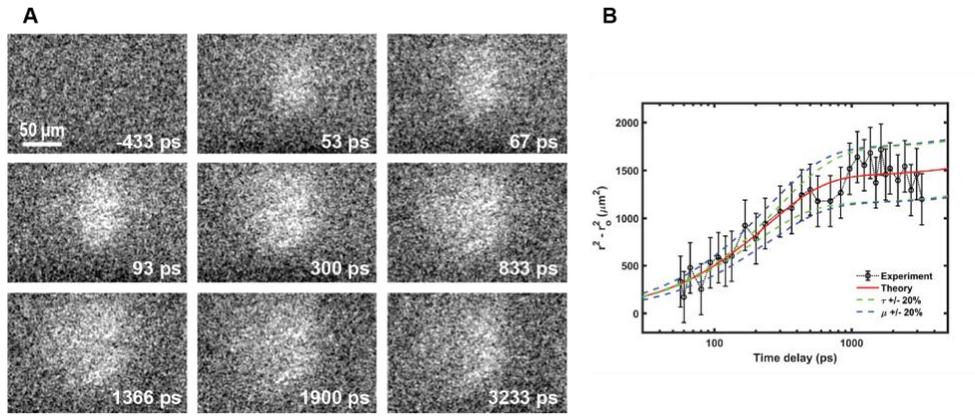

**Fig. S1. Additional SUEM dataset.** Optical fluence: 80 µJ/cm$^2$.

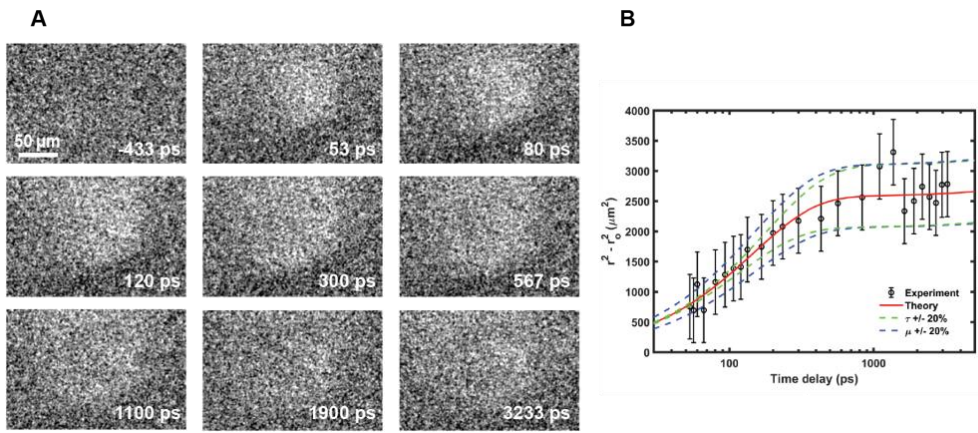

**Fig. S2. Additional SUEM dataset.** Optical fluence: 130 µJ/cm$^2$.

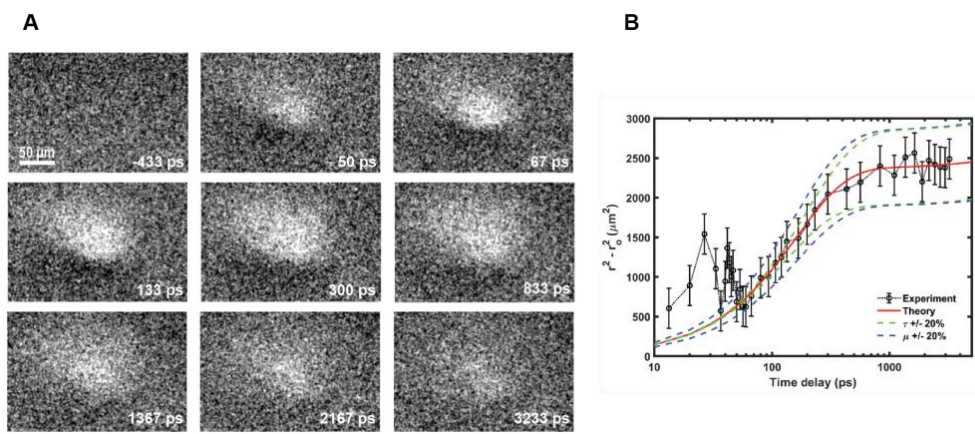

**Fig. S3. Additional SUEM dataset.** Optical fluence: 130 µJ/cm$^2$. High noise in the initial data points is due to weak image contrast before 40 ps.

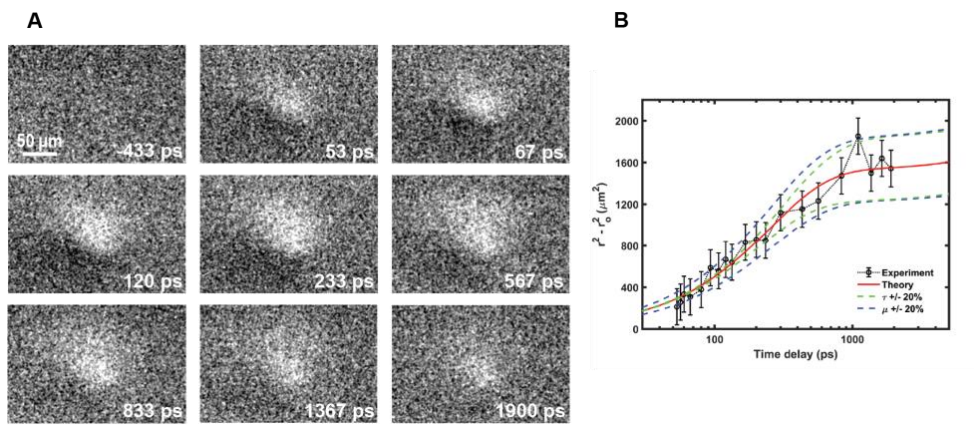

**Fig. S4. Additional SUEM dataset.** Optical fluence: 80 µJ/cm$^2$.

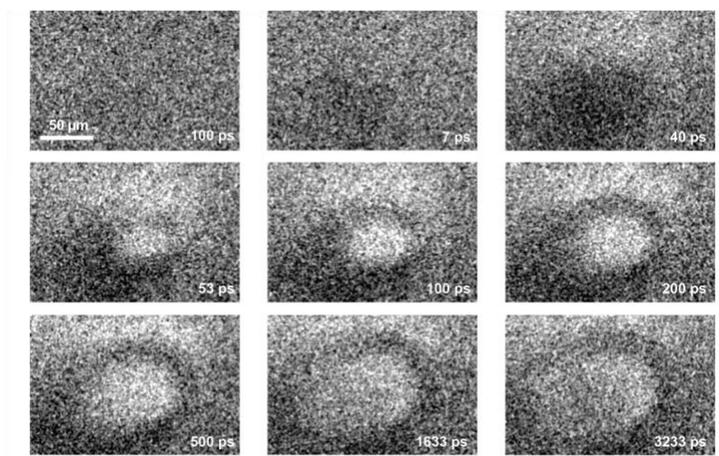

**Fig. S5. Additional SUEM dataset.** Optical fluence: 185 µJ/cm$^2$.

**Movie S1-S7**

Time evolution of the SUEM contrast images for datasets summarized in Table S1.

10